\documentclass[aps,prr,reprint,superscriptaddress]{revtex4-1}

\usepackage{graphicx}
\usepackage{braket}
\usepackage{amsmath,amssymb,amsfonts}
\usepackage{verbatim}
\usepackage{xcolor}

\usepackage{hyperref}

\newcommand{\pref}[1]{(\ref{#1})}
\newcommand{\eref}[1]{Eq.~\pref{#1}}
\newcommand{\fref}[1]{Fig.~\ref{#1}}

\usepackage{tikz}

\begin{document}

\title{Universal shock-wave propagation in one-dimensional Bose fluids}

\author{Romain Dubessy}
\email{romain.dubessy@univ-paris13.fr}
\affiliation{Universit\'e Sorbonne Paris Nord, CNRS UMR 7538, Laboratoire de physique des lasers, F-93430, Villetaneuse, France}

\author{Juan Polo}
\affiliation{Univ. Grenoble Alpes, CNRS, LPMMC, 38000 Grenoble, France}
\affiliation{Quantum Systems Unit, Okinawa Institute of Science and Technology Graduate University, Onna, Okinawa 904-0495, Japan}

\author{H\'el\`ene Perrin}
\affiliation{Universit\'e Sorbonne Paris Nord, CNRS UMR 7538, Laboratoire de physique des lasers, F-93430, Villetaneuse, France}

\author{Anna Minguzzi}
\affiliation{Univ. Grenoble Alpes, CNRS, LPMMC, 38000 Grenoble, France}

\author{Maxim Olshanii}
\affiliation{Department of Physics, University of Massachusetts Boston, Boston, MA 02125, USA}

\date{\today}

\begin{abstract}
We propose a protocol for creating moving, robust dispersive shock waves in interacting one-dimensional Bose fluids.
The fluid is prepared in a moving state by phase imprinting and sent against the walls of a box trap.
We demonstrate that the thus formed shock wave oscillates for several periods and is robust against thermal fluctuations.
We show that this large amplitude dynamics is universal across the whole spectrum of the interatomic interaction strength, 
from weak to strong interactions, and it is fully controlled by the sound velocity inside the fluid.
Our work  provides a generalization of the dispersive shock wave  paradigm to the many-body regime.
The shock waves we propose are within reach for ultracold atom experiments.
\end{abstract}

\maketitle
\section{Introduction}
Large-amplitude moving perturbations are found in all types of fluids, and even in solids. As a response to a sudden change of parameters, a shock wave---a sharp jump in hydrodynamic variables capable of propagating without dispersion---may form. Even ideal fluids can support shock waves as long as the infinitely sharp discontinuities are consistent with the conservation laws. Dissipative effects, present in real-world fluids,  give the shock layer a thickness and a shape~\cite{landau_hydrodynamics}. Superfluids  can host shock waves, within the corresponding hydrodynamic two-fluid theory, as in the case of $^4$He~\cite{Moody1984,Iznankin1983}.  Shock waves were also experimentally observed in dilute, weakly interacting Bose-Einstein condensates of ultracold atoms~\cite{Dutton2001,Simula2005,Chang2008,Meppelink2009,Mossman2018} and fermionic superfluids~\cite{Joseph2011a,Salasnich2011,Ancilotto2012}.

One-dimensional (1D) Bose fluids constitute particularly suitable media for a study of shock waves. Only collective modes are possible in such reduced dimensionality, and the fluids belong to the Luttinger liquid universality class~\cite{haldane1982_569}, thus opening a possibility for a unified theory. Furthermore, at strong interactions, one-dimensional Bose gases display  a statistical transmutation, i.e., some of their properties coincide with those of an ideal Fermi gas, thus allowing for an exact solution~\cite{Girardeau1960}. In addition, several theoretical methods are available in the full spectrum of the interaction strength~\cite{Cazalilla2011,Castro-Alvaredo2016,Bertini2016}, thanks to the integrability of the underlying model~\cite{lieb1963}.

In a strongly interacting 1D Bose gas, shock waves were predicted to form in the time evolution following a density bump in the density profile~\cite{Damski2004}. This protocol creates shock waves that map to a solution breakdown in the nonlinear transport equation (also known as inviscid Burgers' equation): they slowly develop as a result of nonlinearities of the underlying hydrodynamic equations, and then die out after the breakout point.
When applied to a weakly interacting 1D Bose gas, the same protocol also creates shock waves following a similar mechanism~\cite{Damski2004b,Damski2006,Salasnich2015a}. Very recently exact simulations using infinite matrix product states have enabled the study of the dissolution of a density bump through dispersive shock waves (DSW) at intermediate interaction strengths~\cite{Simmons2020}.

\begin{figure}[t]
\includegraphics[width=8cm]{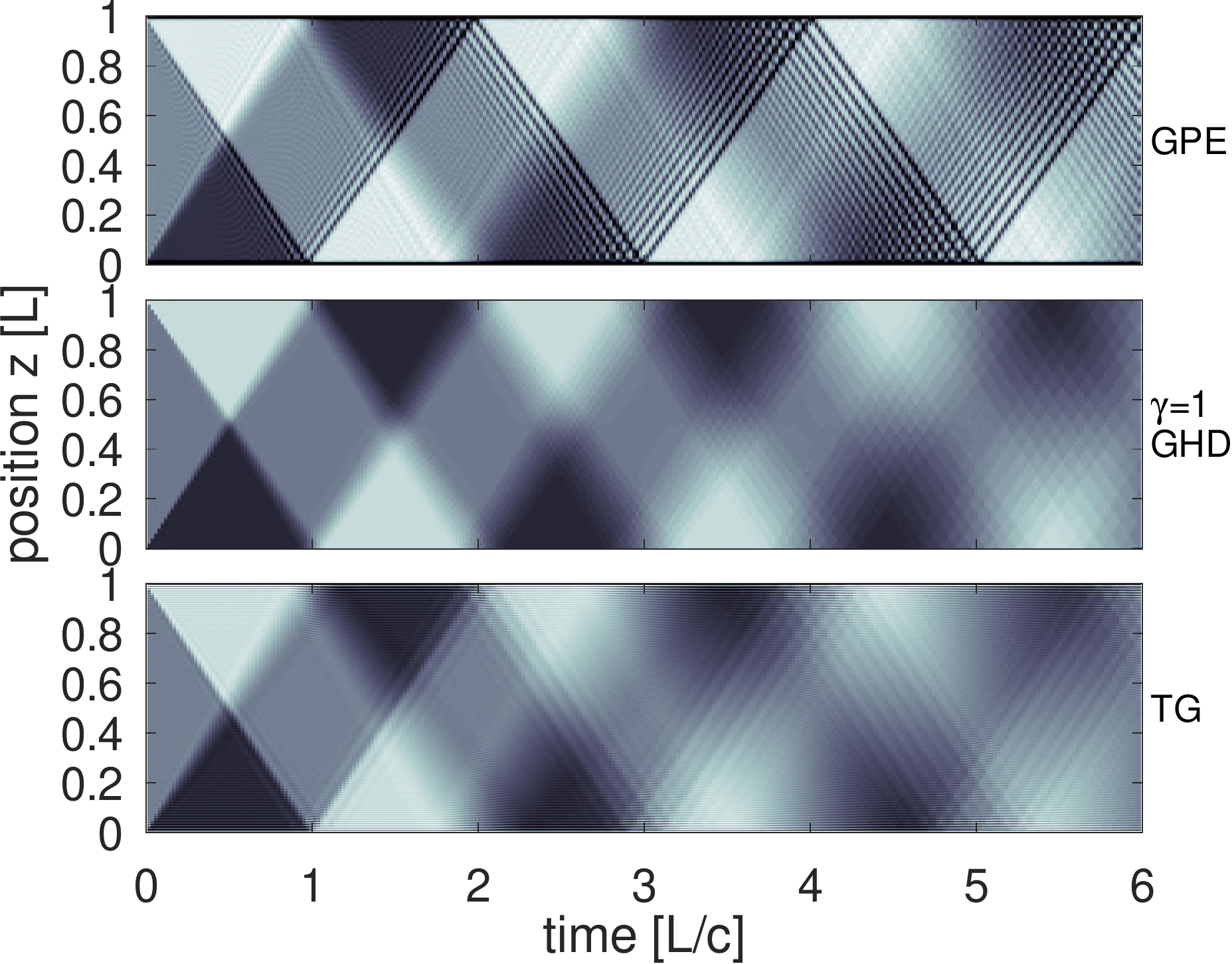}
\caption{\label{fig:density}(Color online) Dynamics of the particle density following a velocity boost of $\sim0.1c(\gamma)$. 
Upper panel: GP regime ($\gamma\ll1$) with $\gamma N^2=2\times10^4$. Middle panel: GHD predictions for $\gamma=1$. Lower panel: TG regime $\gamma\to\infty$, for $N=101$. For each map the time is rescaled by $L/c(\gamma)$.
}
\end{figure}

In our work, we propose a different dynamical protocol for generating propagating shock waves, corresponding to a generalization to the quantum many-body regime of the  combination of a DSW and a rarefaction wave in the mean-field limit (see Ref.~\cite{El2016a} for a review).

By combining three theoretical methods, i.e., classical field theory, generalized hydrodynamics and exact solution we describe all interaction strengths from weak to strong repulsion. The shock-wave front is created when the fluid, with an initially imprinted velocity, hits against the walls of a box trap. Similarly to the solution in the mean-field regime~\cite{El2016a}, the shock wave retains its identity long after it is created, and propagates over several oscillation periods. We observe a remarkably robust behavior of the shock wave propagation at all interaction regimes: we find a universal trend for the wavefront in form of a stable step-like flow and of the current, which displays a triangular-shape oscillation. Both features are robust under inclusion of thermal fluctuations. Our microscopic approaches evidence also non-universal features which depend on the interaction strength: at weak interactions, formation of density modulations due to emissions of phonons and s
 oliton trains, and at large interactions density modulations associated to the Friedel-like oscillations in proximity of a wall, due to quantum fluctuations of the density.

\section{Model}
We consider a one-dimensional Bose gas with repulsive interactions described by the  Lieb-Liniger Hamiltonian:
\begin{equation}
\hat{H}=\int_0^L dz\, \hat{\Psi}^\dagger\left(-\frac{\hbar^2}{2m}\frac{\partial^2}{\partial z^2}+ V(z) +\frac{g}{2}\hat{\Psi}^\dagger\hat{\Psi}\right)\hat{\Psi},
\label{H}
\end{equation}
where $m$ is the mass of the particles, $g$ is the one-dimensional interaction strength~\cite{Olshanii1998} describing the collisions in a tight atomic waveguide. $V(z)$ is a box trap potential of size $L$ with infinitely high walls which we model by imposing hard-wall boundary conditions and $\hat{\Psi}(z)$, $\hat{\Psi}^\dagger(z)$ are bosonic field operators satisfying the commutation relations $[\hat{\Psi}(z),\hat{\Psi}^\dagger(z')]=\delta(z-z')$. The trap contains a fixed number of particles $N=\int_0^L dz\braket{\hat{\Psi}^\dagger\hat{\Psi}}$. We define the dimensionless coupling strength $\gamma=g m/\hbar^2 n_0$, $n_0=N/L$ being the average fluid density.

We study the dynamics following a quench in momentum space: starting from the equilibrium state, at time $t=0$ we apply a phase imprinting to all particles, generated by the shift operator $\hat{U}=e^{i k_0 \hat z}$, yielding a boost of all the particles with velocity $v=\hbar k_0/m$. We follow the quantum dynamics of the particle density $n(z,t)=\langle \hat \Psi^\dagger(z,t) \hat \Psi(z,t)\rangle$ and of the spatial average of the current density $J=- i (\hbar/2m)\int_0^L dz\,\langle \hat{\Psi}^\dagger\partial_z\hat{\Psi}- (\partial_z \hat{\Psi}^\dagger)\hat{\Psi}\rangle$.

The system under consideration is, in general, exactly solvable by Bethe-Ansatz~\cite{Batchelor2005}, however a quench dynamics can be difficult to compute, requiring to evaluate overlaps of excited-state Bethe wavefunctions. Thus, in order to cover the whole interaction range we use three complementary theoretical approaches:  the mean-field Gross-Pitaevskii (GP) equation for the weakly interacting gas~\cite{Gross1961,Pitaevskii1961}, the Generalized Hydrodynamic (GHD) theory for intermediate interactions~\cite{Castro-Alvaredo2016,Bertini2016,Doyon2017} and the time-dependent Bose-Fermi mapping~\cite{Girardeau1960,Girardeau2000a} for the strongly interacting Tonks-Girardeau (TG) limit. We provide here a brief summary of each method and give more details in Appendix~\ref{sec:methods}.

The Gross-Pitaevskii equation\cite{Gross1961,Pitaevskii1961} describes the time evolution of the condensate wavefunction $\psi(z,t)$ \footnote{For $\gamma \ll 1$  we neglect the fluctuations of 1D quasi-condensates and describe the gas by a condensate wavefunction} by the nonlinear Schr\"odinger equation
\begin{equation}\label{GPE}
i\hbar\frac{\partial\psi}{\partial t}=\left(-\frac{\hbar^2}{2m}\frac{\partial^2}{\partial z^2}+V(z)+g\left|\psi\right|^2\right)\psi.
\end{equation}
We solve it numerically by time evolving the initial equilibrium solution $\psi_{\rm eq}(z)$, satisfying the box boundary conditions $\psi_{\rm eq}(0)=\psi_{\rm eq}(L)=0$, boosted by the phase imprinting $\psi(z,t=0)=e^{i k_0 z}\psi_{\rm eq}(z)$.

The Gross-Pitaevskii equation breaks down at intermediate interactions, when quantum fluctuations significantly affect the dynamics and modify the equation of state of the Bose fluid. In this regime, we describe the fluid at long wavelengths using the generalized hydrodynamic equations~\cite{Castro-Alvaredo2016,Bertini2016} for the distribution function $n(z,k,t)$ of the quasiparticles of the Lieb-Liniger model
\begin{equation}
\frac{\partial n}{\partial t}+v_n^{\rm eff}\frac{\partial n}{\partial z}=0,
\label{eqn:GHD}
\end{equation}
solved self-consistently with the equation for the dressed velocity $v_n^{\rm eff}(k)=(\hbar/m)\times([k]^{\rm dr}/[1]^{\rm dr})$,
where the dressing operation is defined by $h^{\rm dr}(k)-\int\frac{dk}{2\pi}\phi(k-k^\prime)n(k^\prime)h^{\rm dr}(k^\prime)=h(k)$ and the  Lieb-Liniger kernel from the Bethe Ansatz solution reads $\phi(k-k^\prime)=2k_c/[k_c^2+(k-k^\prime)^2]$, with $k_c=mg/\hbar^2$ the inverse length scale associated to the interaction strength~\cite{Doyon2017,Bulchandani2017,Doyon2018a}. To implement the quench and impose hard wall boundary conditions we use a mirror image method, see Eq.~\eqref{eqn:boost_fermi_sea}. Once the self-consistent solution $n(z,k,t)$ is found, we compute the current density according to $j(z,t)=\int \frac{dk}{2\pi}\frac{\hbar k}{m}\rho_p(z,k,t)$, where the quasi-particle density $\rho_p\equiv n(z,k,t)\times [1/2\pi]^{\rm dr}$ and the total current  $J(t)=L^{-1}\int_0^L dz\,j(z,t)$.

Finally, in the  Tonks-Girardeau regime of infinitely strongly interacting bosons, we describe the dynamics using an exact solution based on the time-dependent Bose-Fermi mapping~\cite{Girardeau2000a,Wright_2000,Girardeau_2005}, where the many-body wavefunction $\Psi_{TG}(z_1,....,z_N)$ reads
\begin{equation}
\Psi_{TG}(z_1,...,z_N)=\!\!\!\!\!\prod_{1\le \ell'< \ell''\le N}\!\!\!\!\!\text{sgn}(z_{\ell'}-z_{\ell''})\, \left.\text{det}[\psi_{\ell}(z_i,t)]\right|_{\ell,i=1..N},\label{eq:TG_wavefunction}
\end{equation}
where $\psi_\ell(z,t)$ is the solution of the single-particle  Schr\"odinger equation $i \hbar \partial_t \psi_\ell=\left[-\hbar^2 \partial_z^2/2m + V(z)\right]\psi_\ell$ with the initial conditions $\psi_\ell(z,0)=\psi_\ell^{0}(z)$, where $\psi_\ell^{0}(z)$ is the eigenfunction of the Schr\"odinger equation at initial time, boosted by the phase imprinting. This approach allows us to describe in an exact way the full quantum dynamics after the quench.

\begin{figure}[t]
\includegraphics[width=8cm]{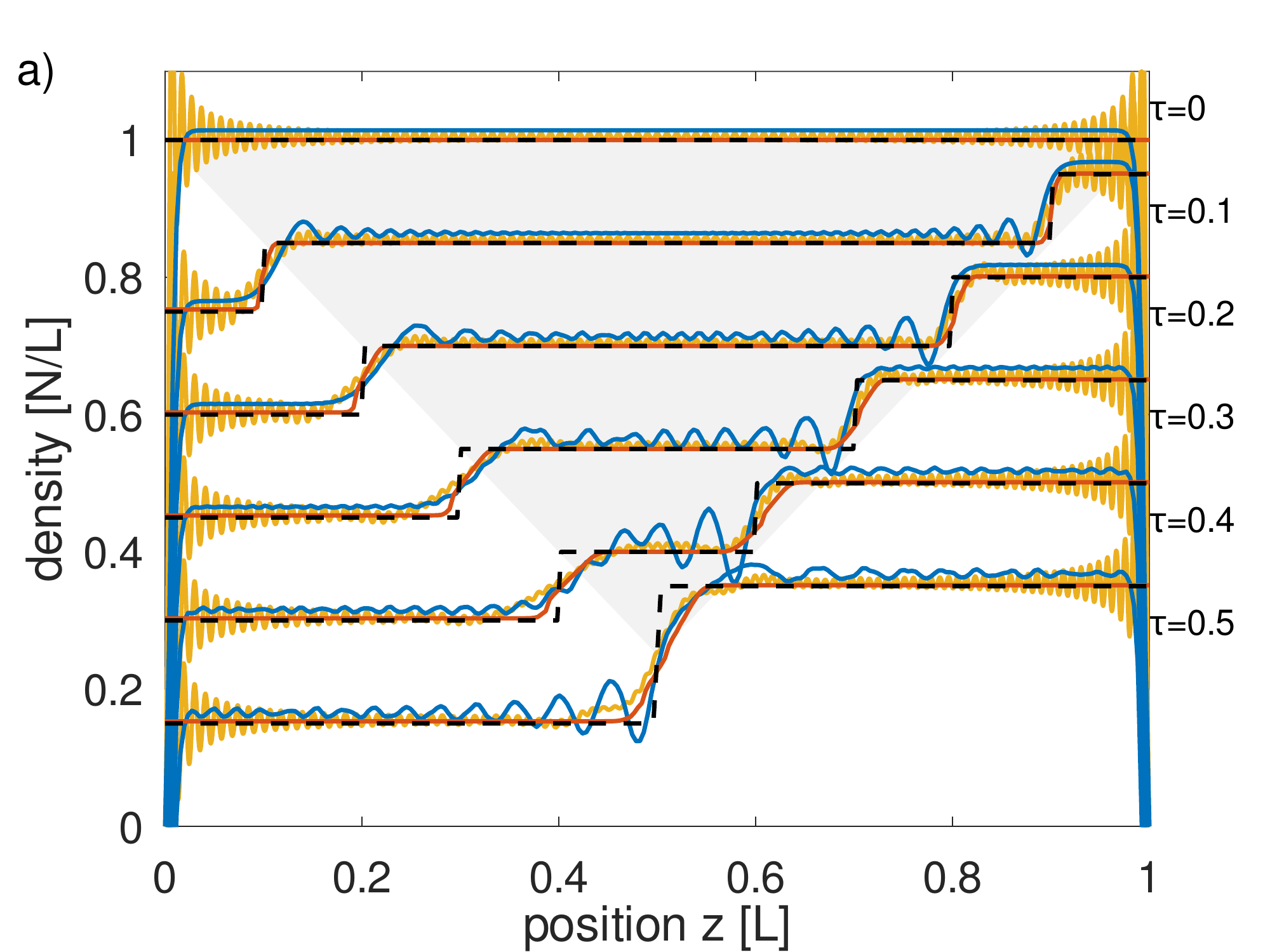}
\caption{\label{fig:density_cut}
(Color online) Particle density $n(z)$ in the box trap (normalized to $N/L$) at different times $t=\tau\times L/c(\gamma)$, using GP (blue solid lines), GHD (orange solid lines) and TG (yellow solid lines) calculations, for the same parameters as in Fig.~\ref{fig:density}.
The dashed black lines are the predictions of the step density profile model (see {\eqref{shock_wave_ansatz} and \eqref{eqn:CHD}}).
The light gray triangle is a guide to the eye emphasizing the common propagation velocity of the fronts.
Particle densities at successive times are shifted downwards by $0.15N/L$ for clarity.
}
\end{figure}

\section{Results}
Figure~\ref{fig:density} shows the universal behavior of the density dynamics. At early times, the density develops a double step profile, corresponding to the shock wave which as we will discuss below generalizes the dispersive shock wave: as the particles are moving towards one side of the box and bounce on the boundary, a high-density plateau develops upstream, while 
a low-density plateau develops downstream as particles move away from the other boundary.
In between the density remains unchanged, until the two plateaus meet and the total current vanishes. At this point the two propagating fronts cross each other (see Fig.~\ref{fig:density_model_GPE_TG})  and the sign of the current is reversed.
Later, the role of the two boundaries being exchanged because the flow is now reversed, the two plateaus develop again, and so on.
As shown in Fig.~\ref{fig:density_cut}, focusing on the early time evolution, the two fronts separating the  density plateaus propagate at the speed of sound $c(\gamma)$~\cite{Lieb1963b} at that fluid density, such that if time is rescaled by $L/c(\gamma)$ the density cuts as a function of time fall onto each other displaying a remarkable universal dynamical behavior. This is even more remarkable once we notice that the density jump occurs on a scale which is of  the order of the healing length of the fluid.

\begin{figure}[ht]
\includegraphics[width=8cm]{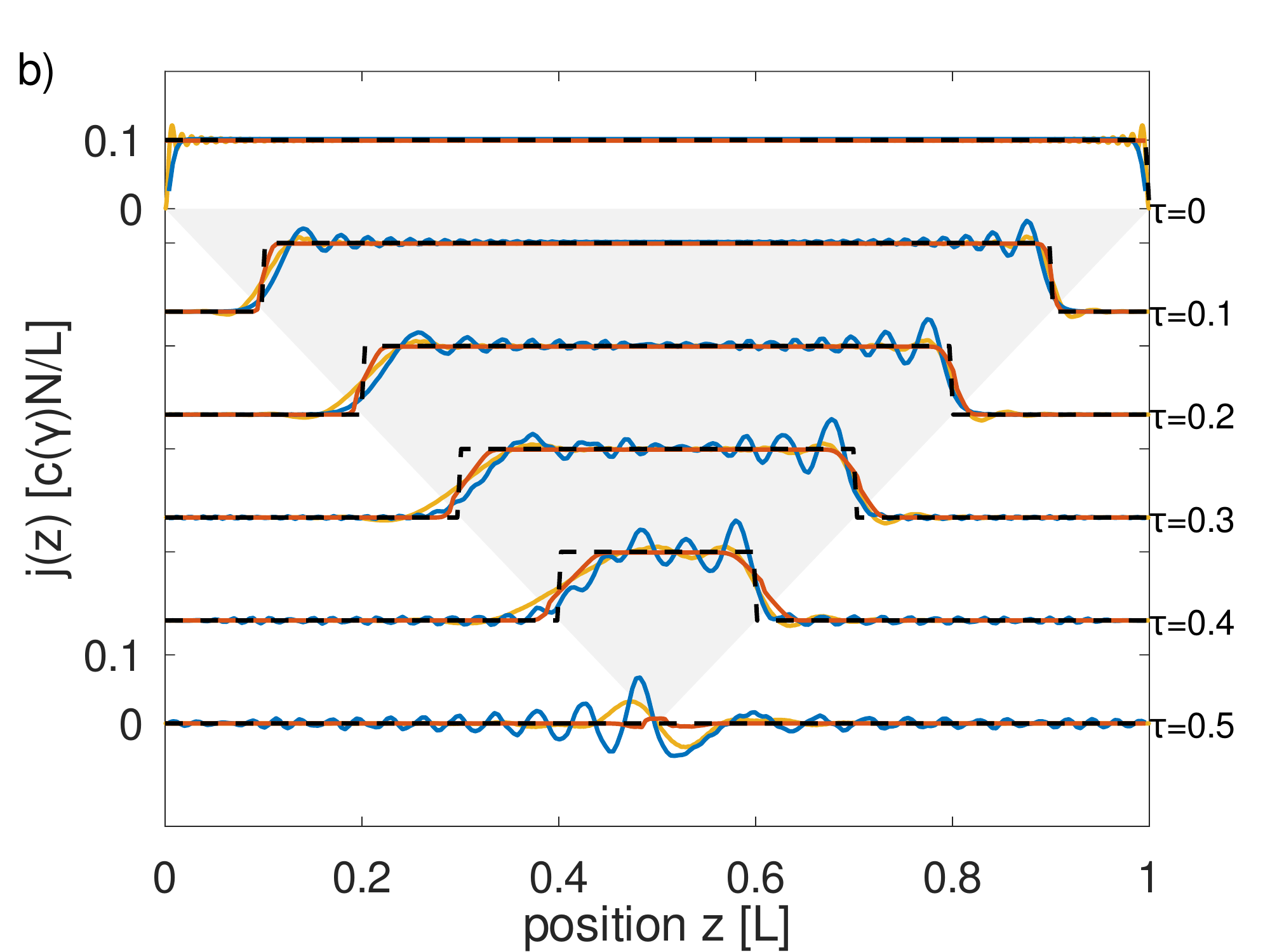}
\caption{\label{fig:localcurrent}
Local current $j(z)$ (normalized to $c(\gamma)N/L$) in
the box potential at different times ($t = \tau \times L/c(\gamma)$), using GP (blue solid lines), GHD (orange solid lines) and TG (yellow solid lines). The dashed black lines are the predictions of the step density profile model. The light gray triangle is a guide for the eye emphasizing the common propagation velocity of the fronts.
Local currents at consecutive times are shifted downwards for clarity.}
\end{figure}

Figure~\ref{fig:localcurrent} shows the local current $j(z)$ as a function of the position coordinate at different times (corresponding to the same times shown in Fig.~\ref{fig:density_cut}). This figure corroborates the universal behavior observed in the density distribution as well as in the total current. Again, we observe that the main features, characterized by the infrared limit, are analogous in all interaction regimes, while the ultraviolet limit is model dependent and presents small deviations between the different curves. 

In addition to universal features, we notice also small differences among the three regimes~\footnote{Note that in the GPE regime, the density bends at the boundaries on a length scale fixed by the healing length, therefore the density plateaus are slightly higher than what is expected in the thermodynamic limit. In the TG solution, the effect of the boundaries is very small on the density plateau level for $N=101$ particles, but we need nonetheless to take into account finite-$N$ corrections in order to estimate the speed of sound accurately~\cite{Batchelor2005}.}:
at weak interactions, in addition to a shock wave, we observe the formation of soliton trains upstream of the flow, see, e.g., Fig.~\ref{fig:density} at times $t=2.5 L/c(\gamma)$, as we have checked by analyzing the phase of the condensate wavefunction, see Fig.~\ref{fig:density_dips}, and also reported in~\cite{Hakim1997,Polo2019a}. At very large interactions, we observe modulations in the density profile, corresponding to Friedel oscillations of the mapped Fermi gas, due to the quantum fluctuations of the density~\cite{haldane1982_569,Didier2009}.

A priori, the quantum many-body Schr\"{o}dinger equation generated by the Hamiltonian \eqref{H} is not guaranteed to support shock waves. Below, we show rigorously how they emerge in the weak interaction limit and suggest why they persist for arbitrary interaction strength.

Small excitations on the surface of flat condensates can be shown to obey a modified Klein-Gordon equation
\begin{equation}
\frac{1}{c_{\rm GP}^2}\frac{\partial^2}{\partial t^2}\delta\psi
-\frac{\partial^2}{\partial z^2}\delta\psi = 
-\frac{1}{4} \xi^2 \frac{\partial^4}{\partial z^4}\delta\psi\,\,,
\label{Klein-Gordon}
\end{equation}
featuring a forth-derivative correction. See Appendix~\ref{sec:methods} and Eqs.~\eqref{Bogoliubov} to \eqref{Bogoliubov3} for a detailed derivation. Here, $c_{\rm GP} \equiv \sqrt{\mu_{0}/m}$ is the speed of sound, and $\xi \equiv \hbar/m c_{\rm GP}$ is the healing length, with $\mu_{0} \equiv g n_{0}$ being the chemical potential before the quench. The correction can be neglected in the long-wave-length limit, and the resulting equation does support features moving at a speed of sound.

Note however, that the equation \eqref{Klein-Gordon} is second order in time, one order higher that the Nonlinear Schr\"{o}dinger equation \eqref{GPE} it was derived from: as such, half of the solutions of \eqref{Klein-Gordon} are spurious, and they should be discarded. Nonetheless, as shown in Appendix~\ref{sec:methods}, it turns out that \eqref{Klein-Gordon}, in the long-wave-length limit $\xi\to 0$, features a moving discontinuity that unites two valid solutions of \eqref{GPE},
\begin{equation}
\psi(z,t) = e^{-i\mu_{0}t/\hbar}
\left\{
\begin{array}{ll}
\sqrt{n_{0}-\Delta n} \, e^{i g \Delta n t/\hbar}
&
z < c_{\rm GP}t
\\
\sqrt{n_{0}} \, e^{i m v z/\hbar - i m v^2 t/2\hbar + i \Delta\phi}
&
z > c_{\rm GP}t
\end{array}
\right.
\,\,,
\label{shock_wave_ansatz}
\end{equation}
provided that the density and velocity discontinuities obey a rigid relationship:
\begin{equation}
\frac{\Delta n}{n_{0}} = \frac{v}{c_{\rm GP}}\, .
\label{eqn:CHD}
\end{equation}

\begin{figure}
\includegraphics[width=8cm]{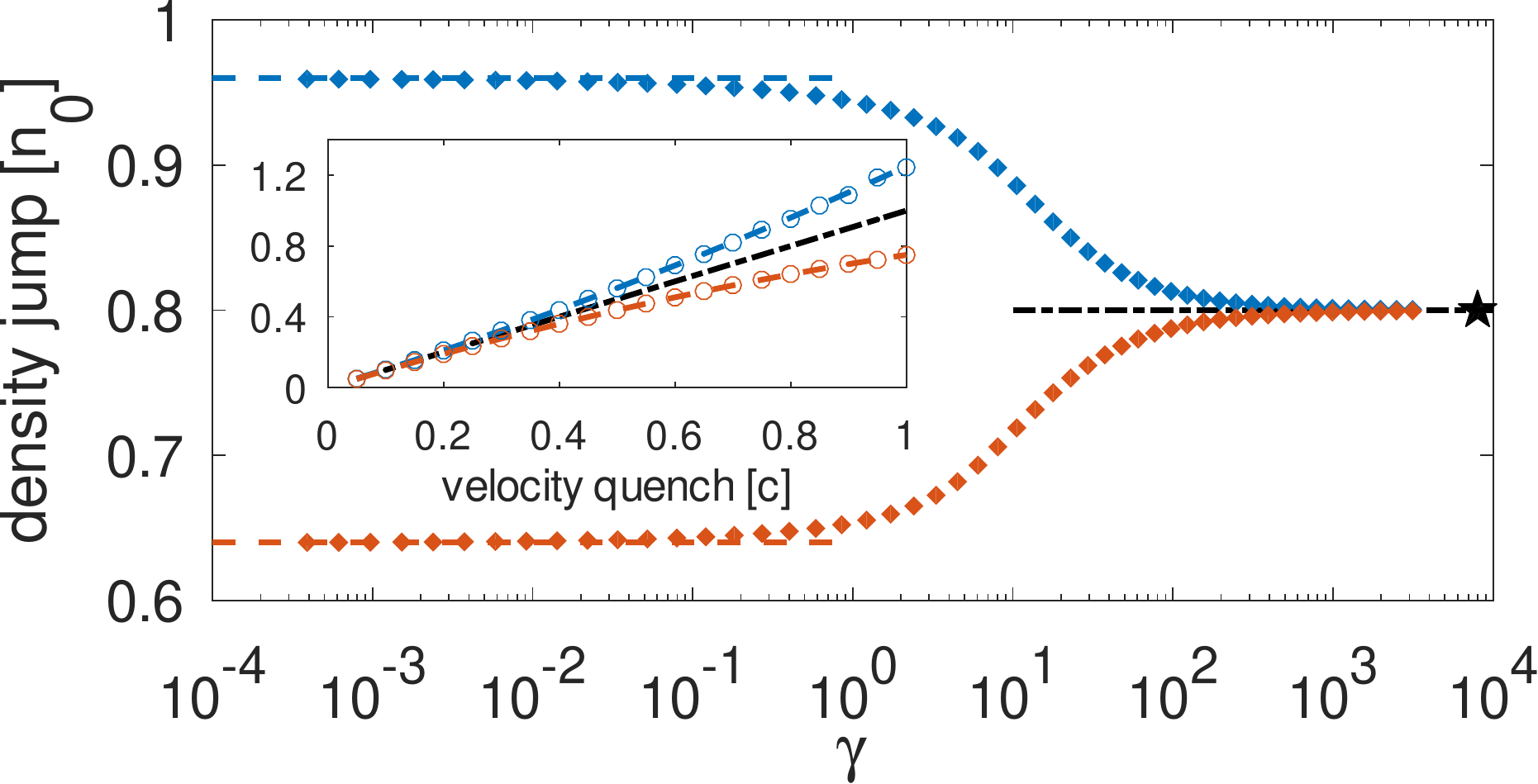}
\caption{\label{fig:jump}
(Color online) Density jumps for a large quench amplitude of $0.8 c(\gamma)$ as a function of $\gamma$  (for full shock wave fronts see Fig.~\ref{fig:GHD_profiles}). Filled diamonds: GHD theory; dashed lines: modulation theory. The blue (orange) color corresponds to higher (lower) jump.  Dot-dashed line: massless Klein-Gordon prediction Eq.~\eqref{eqn:CHD}, with $c_{\text{GP}}$ replaced by $c(\gamma)$.
Filled black star: TG limit. Inset:  density jump in the GP regime as a function of the velocity quench, same color code.
}
\end{figure}

The emergence of Klein-Gordon equation in the infrared limit of the NLS is not an accident: it is rather a manifestation of the bosonization phenomenon~\cite{haldane1982_569}, an emergence of free relativistic bosons, in the long-wavelength limit of one-dimensional systems with phonon excitations, valid in all interaction regimes. Furthermore, for excitations of a macroscopic amplitude, the bosons allow for a classical fields description~\cite{Pedri2008}, yielding \eqref{Klein-Gordon} (with $\xi=0$). The only modification required is that the mean-field speed of sound $c_{\rm GP}$ is replaced by the regime-sensitive, exact speed of sound $c(\gamma)$. Hence, the small discontinuity shock waves should persist in the strongly correlated regimes. Remarkably, this is what we observe numerically, using Eq.~(\ref{eqn:CHD}), with $c_{\rm GP}$ replaced by $c(\gamma)$.

The shape and width of the shock wave front are  regime-specific. Nonetheless, the GHD approximation is able to capture the width of the front in all regimes, agreeing remarkably with the predictions of the Whitham modulation theory~\cite{El2016a} in the GPE regime and with the exact result in the Tonks-Giradeau limit (see the inset of Fig.~\ref{fig:jump} and Fig.~\ref{fig:GHD_profiles}), thereby providing a generalization to the  quantum many body regime of the DSW paradigm.

\begin{figure}
\includegraphics[width=8cm]{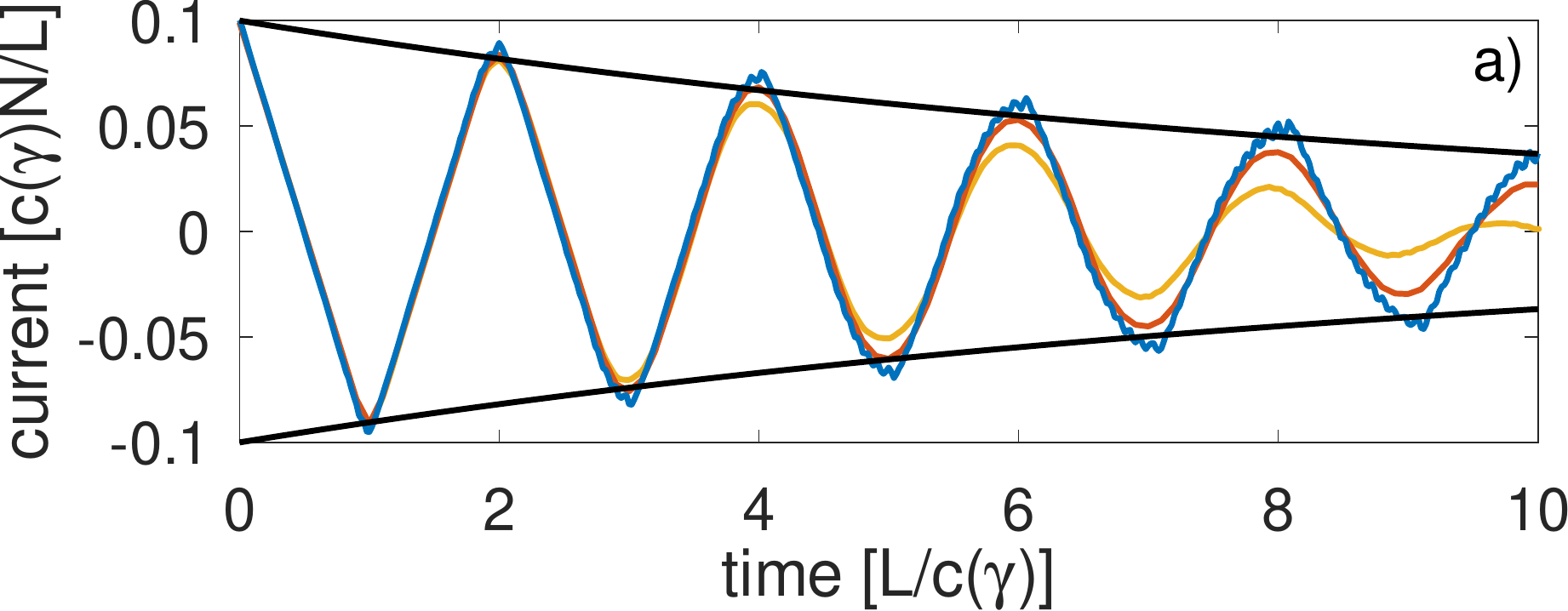}
\includegraphics[width=4cm]{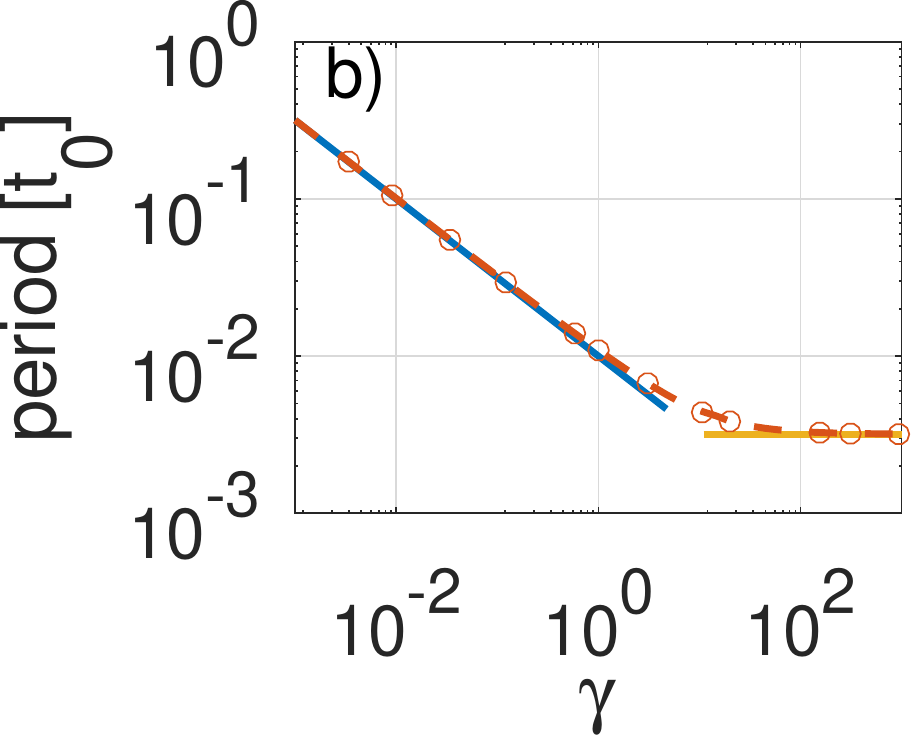}
\includegraphics[width=4cm]{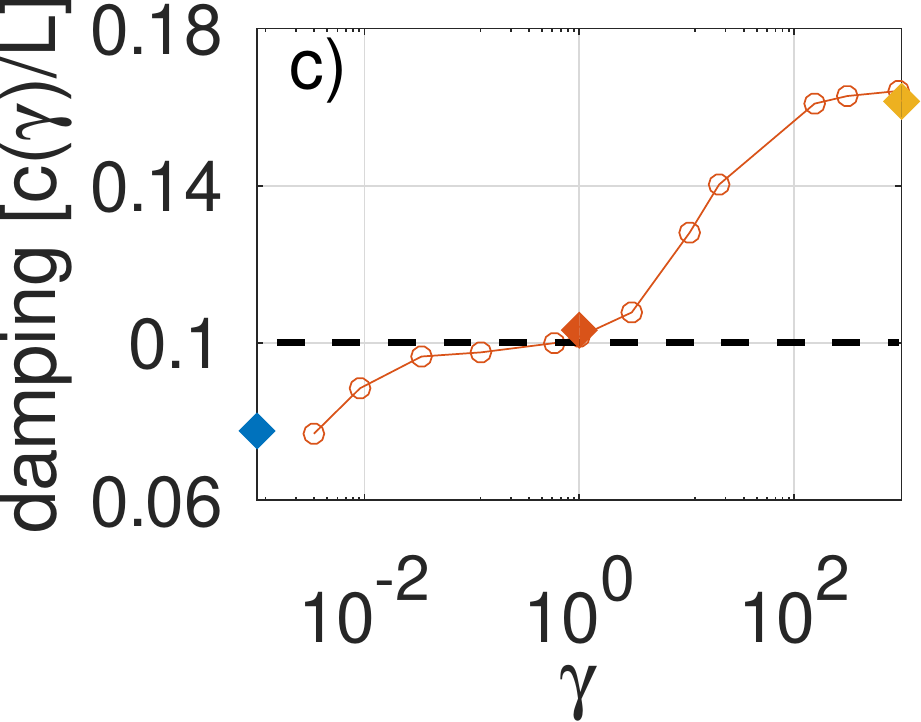}
\caption{\label{fig:current}(Color online) (a) Dynamics of the particle current versus time (in scaled units) for the same parameters as in Fig.~\ref{fig:density}, for GPE (blue), GHD (orange) and TG (yellow) regimes, and their exponential envelope with a time scale $\tau_d$ (black solid lines) (see text). (b-c) Blue dots: (b) period (units: $t_0=mL^2/\hbar$) and (c) damping rate of the current oscillations in a zero--temperature GHD simulation for a quench of amplitude $0.1c(\gamma)$ and $n_0=100$. (b) Dashed black line: ${\mathcal T}=L/c(\gamma)$ (see text). Solid red curves: periods expected in the GPE ($\gamma\ll 1$) and TG ($\gamma\gg 1$) limits. (c) Red diamonds: damping from the data in (a); Horizontal dashed line: inverse of the dephasing time $1/\tau_d$.
}
\end{figure}

Figure~\ref{fig:current} shows the oscillations of the current at longer times obtained from the three theoretical approaches:  GP ($\gamma\ll1$),  GHD at $\gamma=1$, and TG ($\gamma\to\infty$). We have also used  GHD to  investigate the dynamics at the hydrodynamic scale for the whole interaction strength range. We find a good agreement for the current dynamics with the TG exact solution at large $\gamma$ and with the GPE at small $\gamma$. We also obtain the period of the current oscillations, as shown in Fig.~\ref{fig:current}(b). We find that the period is  well accounted for by the expression ${\mathcal T}=L/c(\gamma)$ where $c(\gamma)$ is the exact speed of sound obtained from  the solution of the Lieb-Liniger model~\cite{Lieb1963b}: this provides another confirmation that even though the shock wave is generated by a large-amplitude oscillation, its hydrodynamic nature implies that the speed of sound sets its dynamics.

Our microscopic calculation finally allows us to address the robustness of the shock waves created by the proposed protocol. 
At long times, as illustrated in Fig.~\ref{fig:density} and in Fig.~\ref{fig:density_cut}, the wavefronts  gradually broaden during the propagation, resulting in a loss of contrast between the density plateaus.
Correspondingly, the oscillations of the current  progressively damp and change shape from triangular to sinusoidal, see Fig.~\ref{fig:current}(a).  The damping of the current oscillations weakly increases with interaction strength, and can be estimated within GHD by the dephasing time $\tau_d=L/(v_{\rm high}-v_{\rm low})\sim (n_0/\Delta n)\times L/c(\gamma)$
and hence is faster for stronger quenches (see Fig.~\ref{fig:current}(c)). Here $v_{\rm high}$ ($v_{\rm low}$) corresponds to the effective velocity of the fastest (slowest) quasi-particles involved in the dynamics, as explained in Appendix~\ref{sec:methods}. Its microscopic origin depends on the interaction regime: at weak interactions, it is due to the mode-mode coupling induced by the nonlinearity in the GPE~\cite{Smerzi1997_2}, at strong interactions it is due to the slightly different dispersion of each single-particle mode with time. 

We have also explored the effect of thermal fluctuations in the propagation of the shock waves. We find that the phenomenon persists at finite temperature up to $T \sim \mu/k_B$, with $\mu$ being the chemical potential, and that  the damping of the current oscillations increases with temperature (see Appendix~\ref{app:finiteT} for details). 

\section{Conclusions}
We have proposed a protocol for generating shock waves in a 1D Bose fluid: we use phase imprinting 
to impart a velocity flow onto the gas, driving it against the walls of the container. By combining several theoretical techniques, we have shown that the formed wavefront is stable and propagates over several periods of oscillations in the box trap; 
the effect persists for any interaction strength, from weak to strong repulsion,  and it is robust against thermal fluctuations. We find that even under such a strong quench the wavefront  follows a universal dynamics fixed by the hydrodynamic sound velocity. From the theoretical point of view this means that the underlying microscopic theory supports the universal features and keeps them stable: the large-amplitude dynamics is fully consistent with infrared hydrodynamic regime, and does not depend on short-distance cutoff except for the details of the shape of the wavefront.
 
Our work calls for further studies on the  dynamics at long times, e.g., exploration of the emergence of grey solitons in the weakly interacting regime and their analogues at strong interactions,  and of the origin of the damping mechanisms in one dimension.
More generally, our work constitutes a new avenue towards the theoretical and experimental study of strongly driven one-dimensional quantum systems, allowing for an access to quantum turbulence.
Finally, our result implies an existence of a new kind of universality in out-of-equilibrium dynamics.

%\paragraph{Note added.} After completing this work we became aware of Ref.~\cite{Simmons2020} which studies dispersive shock waves of the type proposed in~\cite{Damski2004}. 

\begin{acknowledgments}
We acknowledge fruitful discussions with J. Dubail on generalized hydrodynamics and the zero entropy subspace method.
JP acknowledges Okinawa Institute of Science and Technology Graduate University and also the JSPS KAKENHI Grant Number 20K14417. We acknowledge financial support from the ANR project SuperRing (Grant No. ANR-15-CE30-0012). LPL is a member of DIM SIRTEQ (Science et Ing\'enierie en R\'egion \^Ile-de-France pour les Technologies Quantiques). 
MO acknowledges support from the National Science Foundation grants PHY-1912542 and PHY-1607221.
\end{acknowledgments}

\appendix

\section{Methods}
\label{sec:methods}

Here we provide details on the different methods and approaches used in the main text.

\paragraph{\bf{Details on the solution of the Gross-Pitaevskii equation}}
To describe the dynamics in the  $\gamma\ll 1$ regime we solve the Gross-Pitaevskii equation \eqref{GPE} numerically, using a spectral method relying on the discrete sine transform embedding the hard wall boundary conditions $\psi(0)=\psi(L)=0$.
We first use imaginary time propagation to find the ground state in the box, then we quench the state at $t=0$ and compute the subsequent dynamics. To ensure that the system is in the mean-field hydrodynamic regime we choose a sufficiently large non linear coefficient $\gamma N^2=gN\times mL/\hbar^2=20000$. We have checked that the transition between the single particle and mean-field regime occurs at $\gamma N^2\sim 500$.

\paragraph{\bf{Details on the solution of the generalized hydrodynamics equations}}
The main GHD equation is given in Eq.~\eqref{eqn:GHD}, which we recall here: 
\[
\frac{\partial n}{\partial t}+v_n^{\rm eff}\frac{\partial n}{\partial z}=0,
\]
where $n$ is the occupation function of the Lieb-Liniger quasi-particles, and the dressed velocity is given by:
\[
v_n^{\rm eff}(k)=\frac{\hbar}{m}\frac{[k]^{\rm dr}}{[1]^{\rm dr}}.
\]
At first the GHD formalism seems incompatible with the box boundary conditions, because it relies on the local density approximation. One method to naturally include the effect of the hard-wall boundaries is to double the system size (from $[0,L]$ to $[-L,L]$), impose periodic boundary conditions with period $2L$, and use an anti-symmetric initial state: the right part $z\geq0$ (resp. left part $z<0$) is quenched with a positive (resp. negative) velocity boost: 
\begin{equation}\label{eqn:boost_fermi_sea}
n_0(z,k)=\begin{cases}
\bar{n}(k-k_0) & z\geq 0,\\
\bar{n}(k+k_0) & z < 0,
\end{cases}
\end{equation}
where $\bar{n}(k)$ is the equilibrium occupation function obtained from the equation of state.
This approach is well adapted to GHD and exact at the level of the initial Lieb-Liniger Hamiltonian.

To integrate the GHD equations at zero temperature we use the zero entropy subspace method~\cite{Doyon2018a}.
In this case it is sufficient to compute the evolution of the edges of the Fermi sea, that are located initially at $k=\pm K$~\cite{lieb1963}. After the quench described by~\eqref{eqn:boost_fermi_sea}, the edges are shifted to $\pm K+ k_0$. Furthermore the box boundary condition imposes that a quasi-particle arriving at the right boundary with quasi-momentum $k>0$ is reflected at quasi-momentum $-k$ (particles at $k<0$ are already moving away from the boundary). A symmetric condition occurs at the left boundary. Therefore, immediately after the quench, the dynamics of the front moving to the left is fixed by the quasi-particles lying in $k\in[-K-k_0,-K+k_0]$, while the front moving to the right corresponds to quasi-particles in $k\in[K-k_0,K+k_0]$. The broadening of the fronts is then explained by the fact that these quasi-particles move at different effective velocities: for example, the width of the front moving to the right will evolve as: $t\times(v_{\rm high}-v_{\rm low}
 )$, where $v_{\rm high}=v_n^{\rm eff}(K+k_0)$ and $v_{\rm low}=v_n^{\rm eff}(K-k_0)$.

\begin{figure}[t]
\begin{tikzpicture}
\node at (-3.7,1.75) {a)};
\fill[black!20] (-3.5,-1.5) -- ++(0.5,0) -- ++(0,3) -- ++(-0.5,0) --cycle;
\fill[black!20] (3,-1.5) -- ++(0.5,0) -- ++(0,3) -- ++(-0.5,0) --cycle;
\draw[ultra thick] (-3,-1.5) -- (-3,1.5) node[above] {$x=0$} (3,-1.5) -- (3,1.5) node[above] {$x=L$};
\draw[->] (-3.7,0) -- (3.7,0) node[above] {$x$};
\draw[->] (0,-1.5) -- (0,1.5) node[left] {$k$};
\draw[dashed] (-3,-0.9) node[left] {$-K$} -- ++(0,1.8) node[left] {$K$} -- ++(6,0) -- ++(0,-1.8) --cycle;
\draw[blue,fill=blue!10,fill opacity=0.3] (-3,-0.9+0.3) -- ++(0,1.8) -- ++(6,0) -- ++(0,-1.8) --cycle;
\draw[->,blue] (-1,-0.9) -- node[right] {$k_0$} ++(0,0.3);
\end{tikzpicture}
\begin{tikzpicture}
\node at (-3.7,1.75) {b)};
\fill[black!20] (-3.5,-1.5) -- ++(0.5,0) -- ++(0,3) -- ++(-0.5,0) --cycle;
\fill[black!20] (3,-1.5) -- ++(0.5,0) -- ++(0,3) -- ++(-0.5,0) --cycle;
\draw[ultra thick] (-3,-1.5) -- (-3,1.5) (3,-1.5) -- (3,1.5);
\draw[->] (-3.7,0) -- (3.7,0) node[above] {$x$};
\draw[->] (0,-1.5) -- (0,1.5) node[left] {$k$};
\draw[dashed] (-3,-0.9) -- ++(0,1.8) -- ++(6,0) -- ++(0,-1.8) --cycle;
\draw[blue,fill=blue!10,fill opacity=0.3] 
	(-3,-0.9+0.3) -- ++(0,1.8-0.6) -- ++(1.5,0) node[below,opacity=1] {$A$}
	-- ++(0.5,0.6) node[above,opacity=1] {$B$} -- ++(6-2,0) -- ++(0,-1.8-0.6)
	-- ++(-2,0) node[below,opacity=1] {$C$} -- ++(0.5,0.6) node[above,opacity=1] {$D$}
	--cycle;
\end{tikzpicture}
\begin{tikzpicture}
\node at (-3.7,1.75) {c)};
\fill[black!20] (-3.5,-0.25) -- ++(0.5,0) -- ++(0,1.75) -- ++(-0.5,0) --cycle;
\fill[black!20] (3,-0.25) -- ++(0.5,0) -- ++(0,1.75) -- ++(-0.5,0) --cycle;
\draw[ultra thick] (-3,-0.25) -- (-3,1.5) (3,-0.25) -- (3,1.5);
\draw[->] (-3.7,0) -- (3.7,0) node[above] {$x$};
\draw[->] (0,-0.25) -- (0,1.5) node[left] {$\rho$};
\fill[blue!10,opacity=0.3] (-3,0.8) -- ++(1.5,0) -- ++(0.5,0.2) -- ++(2,0) -- ++(0.5,0.2) -- ++(1.5,0)
							-- ++(0,-1.2) -- ++(-6,0) -- cycle;
\draw[thick,blue] (-3,0.8) -- ++(1.5,0) -- ++(0.5,0.2) -- ++(2,0) -- ++(0.5,0.2) -- ++(1.5,0);
\end{tikzpicture}
\caption{\label{fig:GHD_sketch}
Sketch of the GHD dynamics at $T=0$: the light blue shaded shape indicates the area where $n(z,k,t)=1$, inside the box potential. a) State just after the quench: boosted Fermi sea. b) After a time $t<L/(2c)$ the occupation function $n(z,k,t)$ acquires a non trivial structure and the dynamics is mainly encoded in the position of the points $A$, $B$, $C$ and $D$. c) Sketch of the real space density $\rho(z)$ corresponding to the state of b).}
\end{figure}
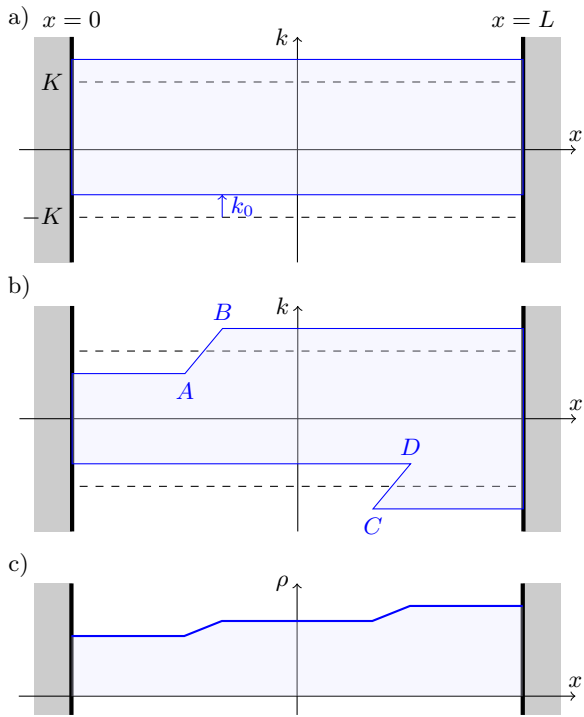

\begin{figure}[t]
\includegraphics[width=8cm]{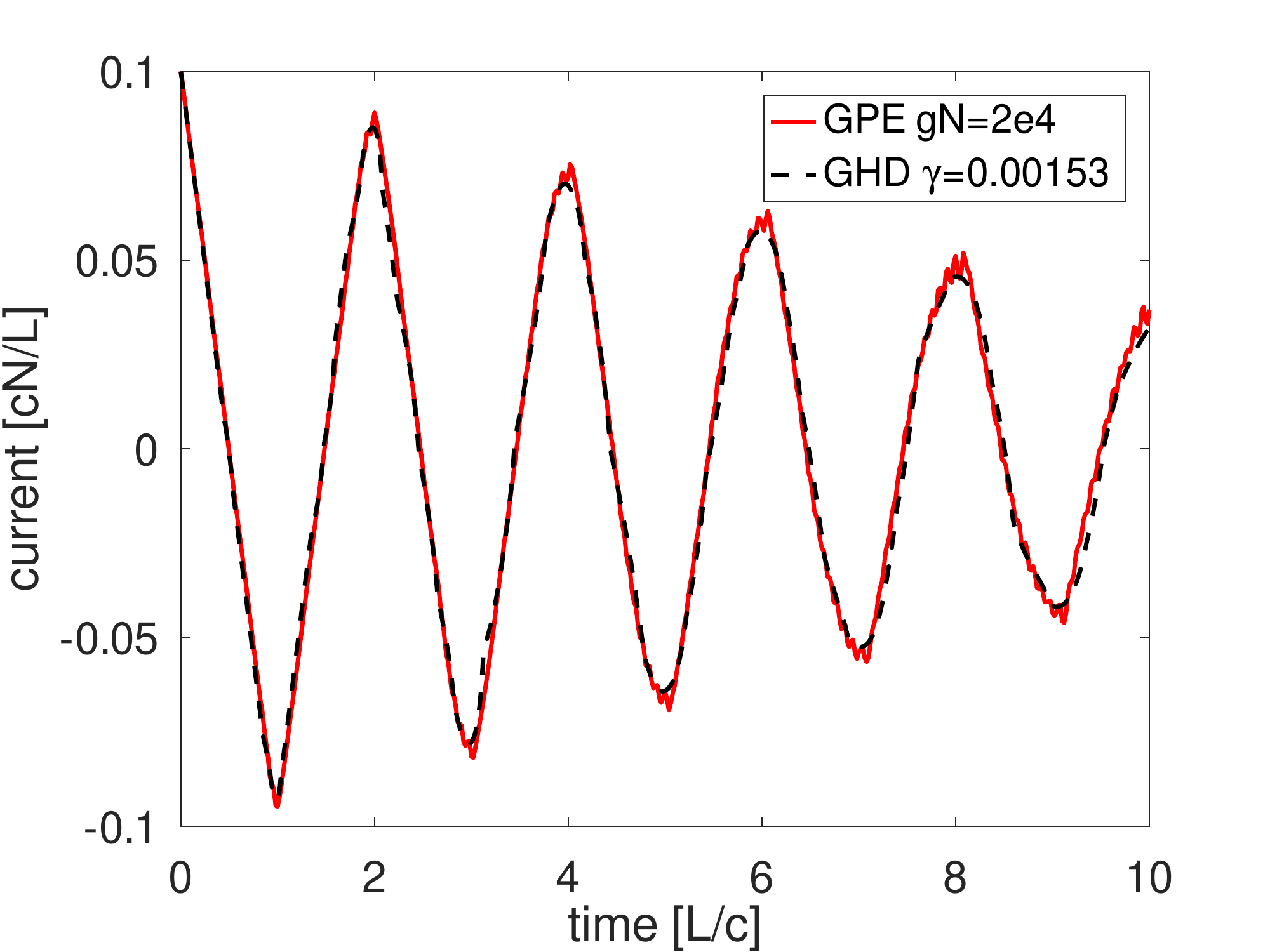}
\caption{\label{fig:GPE}Current as a function of time using GP and GHD at small $\gamma$. The system is quenched with a velocity boost of $0.1c(\gamma)$.}
\end{figure}

\begin{figure}[t]
\includegraphics[width=8cm]{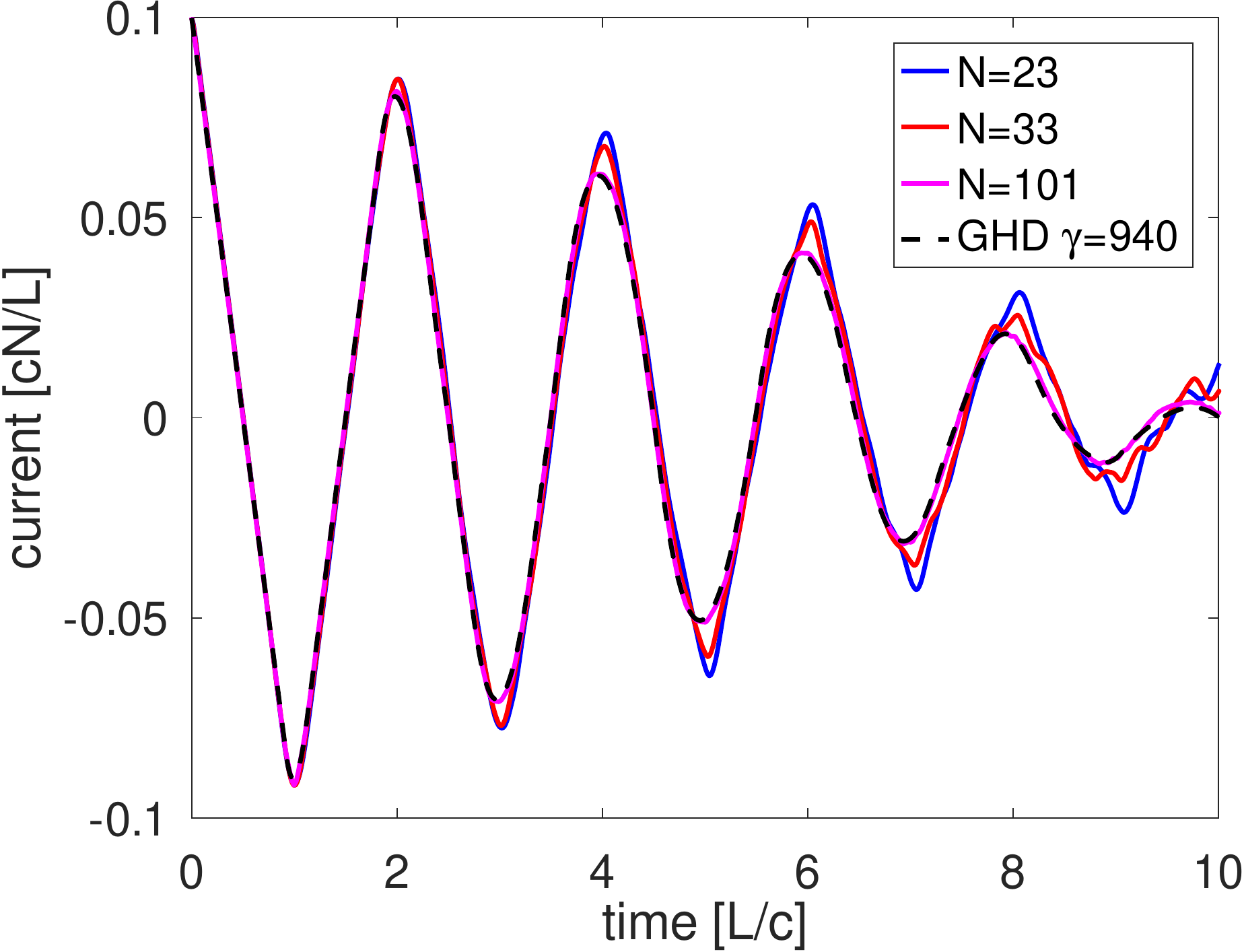}
\caption{\label{fig:tonks}Current as a function of time using the exact TG solution and the GHD,
at large $\gamma$. 
The system is quenched with a velocity boost of $0.1c(\gamma)$.
The $N=101$ TG and GHD curves are undistinguishable at this scale.
}
\end{figure}

This simple explanation indicates that both fronts broaden within GHD, as is seen in the simulation and sketched in Fig.~\ref{fig:GHD_sketch}. Therefore, GHD is not able to reproduce the microscopic details of the exact GP and TG results, while giving an accurate prediction for global observables, see figures~\ref{fig:GPE} and \ref{fig:tonks}.
However it is interesting to notice that for small quenches the shock wave front corresponds in the GHD solution to a local Fermi sea with a hole --see in Fig.~\ref{fig:GHD_sketch}b) the structure between points $C$ and $D$--, as `hole states' in the Lieb-Linieger model give rise to the celebrated Lieb-II spectrum~\cite{Lieb1963b}, often interpreted in the mean-field limit as a solitonic branch. Within this picture the GP and GHD models agree: the shock front resolves through soliton--like excitations.  The same picture holds in the TG regime where the Lieb-II branch corresponds to delocalized solitons, which hence, at different from the GP regime,  cannot be resolved.

We have also  benchmarked our results using an independent integration scheme, based on the iterative method of~\cite{Bulchandani2017}, which also allows for finite temperature calculations.
To summarize, the occupation function at time $t+dt$ is obtained by solving the implicit equation:
\begin{equation}
n(z,k,t+dt)=n\left(z-v^{\rm eff}_{n(z,k,t+dt)}dt,k,t\right).
\label{eqn:implicit_GHD}
\end{equation}
This is done by iterating this formula starting with the initial guess $n(z,k,t+dt)=n(z,k,t)$. During this process, periodic boundary conditions are enforced on the interval $[-L,L]$. To proceed numerically, we use a discrete rectangular grid to store the values of $n(z,k,t)$ at time $t$ and rely on a cubic interpolation formula on this grid to evaluate equation~\eqref{eqn:implicit_GHD}. Once satisfactory convergence is obtained the same method is repeated to compute the next time step, until the desired final time is achieved.

\paragraph{\bf{Details  on the Tonks-Girardeau exact solution} }

In the infinitely strongly repulsive limit, $\gamma\rightarrow\infty$, we focus on the exact Tonks-Girardeau (TG) solution~\cite{Girardeau1960}. In particular, we make use of the time-dependent Bose-Fermi mapping~\cite{Girardeau2000a,Wright_2000,Girardeau_2005}, where the many-body wavefunction $\Psi_{TG}$ is written in Eq.~\eqref{eq:TG_wavefunction}. 

Our specific protocol is the following: we write the initial wavefunction as the ground state of a hard-wall box potential, constructed by the first $N$ single-particle orbitals $\chi_\ell(z)$, which we then multiply by a phase profile, induced by the phase imprinting, obtaining the wavefunction $\psi_\ell^{0}(z)=e^{i k_0 z}\chi_\ell(z)$, which is used as starting point for the time evolution. The evolution is then calculated by projecting this state in the eigenbasis of the unperturbed system $\psi_{\ell'}(z,t)=\sum_{\ell}^{\infty}\langle\chi_\ell|\psi_{\ell'}^{0}\rangle \chi_\ell(z)e^{-i\epsilon_{\ell} t/\hbar}$ and where $\epsilon_{\ell}$ is the $\ell$-th single-particle eigenenergy~\cite{Millard_1969,Wright_2002}.

The current of a TG gas at finite temperature is then readily obtained in terms of the evolved single-particle orbitals according to
\begin{equation}
j(z,t)=\frac{\hbar}{m}\text{Im}\left[\sum_{\ell}^{\infty}f(\epsilon_\ell)\psi_\ell^{*}(z,t)\partial_z\psi_\ell(z,t)\right] \label{eq:current_temp_TG}
\end{equation}
with $f(\epsilon)$ being the Fermi-Dirac distribution.
In our specific quench setup, the current density after the phase imprinting reads
\begin{align}
j(z,t)=&\frac{\hbar}{Nm}\text{Im}\bigg[\sum_{\ell}^{\infty}\sum_{\ell'}^{\infty} A_{\ell',\ell}(z)  e^{-i(\epsilon_{\ell'} - \epsilon_\ell)t/\hbar} \bigg],
 \label{eq:current_temp_TG_full}
\end{align}
with an  amplitude of the excitations being given by 
\begin{align}
A_{\ell,\ell'}(z) =& \frac{\hbar}{mL}\text{Im}\bigg[\sum_{i}^{\infty}f(\epsilon_i)\langle\chi_i|\psi_{\ell'}\rangle \langle\psi_{\ell}|\chi_i\rangle \psi_{\ell'}^{*}(z)\partial_z\psi_{\ell}(z) \bigg].
\label{eq:ampltude_excittaion}
\end{align}

The sound velocity of a TG gas is readily obtained from its equation of state. In order to compare it with the  generalized hydrodynamics predictions, we have included the first order correction due to the boundary~\cite{Batchelor2005}, such that at zero temperature it reads:
\[
c_{\rm TG}=\frac{\hbar\pi n}{m}\sqrt{1+\frac{3}{2N}}.
\]
This correction has been included in all figures appearing in the main paper. It is particularly relevant to obtain the proper rescaling of the time axis, leading to an almost perfect collapse of all density and current dynamics curves obtained with the three different approaches considered here.
In addition to this correction, finite size effects can also play an important role in the long time dynamics, as the TG time evolution exhibits revivals at $T_r=N c_{\rm TG}/L$.

\paragraph{\bf{Details of the derivation of the Klein-Gordon form of the Bogoliubov equations}}  
Consider the Gross-Pitaevskii equation \eqref{GPE}, and set the potential $V(z)$ to zero.
Assuming a weak perturbation on top of a density $n_0$ and neglecting box boundaries we set $\psi(z,t) = \left(\sqrt{n_{0}} + \delta\psi(z,t)\right)e^{-i \mu_{0}t/\hbar},$ with $\mu_{0} \equiv g n_{0}$. To the first order in $\delta\psi$, this field itself obeys 
\begin{align}
i\hbar\frac{\partial}{\partial t}\delta\psi = -\frac{\hbar^2}{2m} \frac{\partial^2}{\partial z^2} \delta\psi + \mu_{0} \delta\psi + \mu_{0}\delta\psi^{\star}
\,\,.
\label{Bogoliubov}
\end{align}
Our goal is to eliminate the complex conjugate field. To that end, we can differentiate 
\eqref{Bogoliubov} with respect to time and obtain: 
\begin{align}
\frac{m}{\hbar}\frac{\partial^2}{\partial t^2}\delta\psi
=
\frac{i}{2}\frac{\partial^3}{\partial z^2 \partial t}\delta\psi 
-
\frac{i}{\xi^2} \frac{\partial}{\partial t}\delta\psi 
-
\frac{i}{\xi^2} \frac{\partial}{\partial t}\delta\psi^{\star} 
\,\,,
\label{Bogoliubov2}
\end{align}
with $\xi \equiv \hbar/m c_{\rm GP}$ is the healing length. In the 
r.h.s., the first 
time derivatives and $\delta\psi^{\star}$ itself can be eliminated using 
\eqref{Bogoliubov} and its complex conjugate, yielding 
\begin{equation}
\frac{1}{c_{\rm GP}^2}\frac{\partial^2}{\partial t^2}\delta\psi
-\frac{\partial^2}{\partial z^2}\delta\psi = 
-\frac{1}{4} \xi^2 \frac{\partial^4}{\partial z^4}\delta\psi\,\,.
\label{Bogoliubov3}
\end{equation}

\paragraph{\bf{Details of the derivation of the constraints on the discontinuities across the shock wave front imposed by the consistency
between the GPE equation and the wave equation}  }
Not every solution of the wave equation \eqref{Klein-Gordon} is a proper low-amplitude long-wavelength limit of a solution of the 
Gross-Pitaevskii equation \eqref{GPE}, but some are. Consider a such limit for the Ansatz \eqref{eqn:CHD}.  Assuming the density depression $\Delta n$, the velocity $v$, and the phase $\Delta\phi$ be small (with the smallness of $v$ being required in the long wavelength
limit) and of the same order in variation of the base solution of the Schr\"{o}dinger equation, we can expand the expression \eqref{eqn:CHD} to the first order in $\Delta n$, $v$, and $\Delta \phi$, arriving at 
\begin{align*}
\delta\psi(z,t) = \sqrt{n_0}\times
\left\{
\begin{array}{ll}
-\frac{1}{2}\frac{\Delta n}{n_{0}} + i g\Delta n t/\hbar
&
z < c_{\rm GP}t
\\
i m v z/\hbar + i \Delta \phi
&
z > c_{\rm GP}t
\end{array}
\right\}
\,\,.
\end{align*}
Notice that this expression is not, a priori, in the required form :
\[
\delta\psi(z,t)=\sqrt{n_0}\left[f_+(z-c_{\rm GP}t)+f_-(z-c_{\rm CG}t)+f_0\right]
\]
a solution of a wave equation must yield. However, if we impose
\begin{align}
\frac{\Delta n}{n_{0}} = \frac{v}{c_{\rm GP}}
\,\,,
\label{relationship_between_discontinuities}
\end{align}
the fields $f_+$ and $f_-$ become readily available:
\begin{align*}
&
f_+(\zeta) = \frac{imv}{\hbar}\zeta \Theta[\zeta] - \frac{imv}{2\hbar}\zeta + \left(\frac{v}{2c_{\rm GP}} + i\Delta\phi\right)\Theta[\zeta] 
\\
&
f_-(\zeta) = +\frac{imv}{2\hbar}\zeta
\\
&
f_0 = -\frac{v}{2c_{\rm GP}}
\,\,,
\end{align*}
leading to
\begin{align}
\begin{split}
&
\delta\psi(z,t) = \sqrt{n_{0}}
\Big(
\frac{imv}{\hbar}(z-c_{\rm GP}t)\Theta[z-c_{\rm GP}t]  
+
\frac{imv}{\hbar} c_{\rm GP}t
\\
&
\qquad
-
\frac{v}{2c_{\rm GP}}\Theta[-(z-c_{\rm GP}t)] + i\Delta\phi\, \Theta[z-c_{\rm GP}t]
\Big)\,.
\end{split}
\label{shock_wave_in_Luttinger}
\end{align}
Note that (a) it can be shown that the relationship \eref{relationship_between_discontinuities} is fully consistent 
with---and is, in fact, necessary for---conservation of matter;
(b) the small phase jump $\Delta \phi$ remains undetermined. 

%\section{Local current}
%We complement Fig.~\ref{fig:density_cut} and Fig.~\ref{fig:jump} by showing the same universal behavior in the local current when using the three different approaches considered in this work GPE, GHD and TG.
%Figure~\ref{fig:localcurrent} shows the local current $j(z)$ as a function of the position coordinate at different times (corresponding to the same times shown in Fig.~\ref{fig:density_cut}). This figure corroborates the universal behavior observed in the density distribution as well as in the total current. Again, we observe that the main features, characterized by the infrared limit, are analogous in all interaction regimes, while the ultraviolet limit is model dependent and presents small deviations between the different curves. 
%\begin{figure}[ht]
%\includegraphics[width=8cm]{FigS2}
%\caption{\label{fig:localcurrent}
%Local current $j(z)$ (normalized to $c(\gamma)N/L$) in the box potential at different times ($t = \tau \times L/c(\gamma)$), using GP (blue solid lines), GHD (orange solid lines) and TG (yellow solid lines). The dashed black lines are the predictions of the step density profile model. The light gray triangle is a guide for the eye emphasizing the common propagation velocity of the fronts. Local currents at consecutive times are shifted downwards for clarity.}
%\end{figure}

\section{Benchmark of generalized hydrodynamic predictions at weak and strong interactions}

Figure~\ref{fig:GPE} shows a comparison of the  dynamics of the particle current at weak interactions,  according to   the predictions of the Gross-Pitevskii equation  and of the GHD solution at $\gamma=1.5\times10^{-3}$. The agreement is very good, both for the oscillation frequency and the decay time.

Figure~\ref{fig:tonks} compares the current dynamics from the exact Tonks-Girardeau result for $N=23$, $N=33$, and $N=101$, and the GHD simulation at $\gamma=940$.
The GHD and exact Tonks-Girardeau solution agree very well for $N=101$, thereby benchmarking the validity of the GHD predictions also at strong interactions. At lower number of particles we attribute the discrepancies to finite size effects, that are not captured within GHD. Our analysis shows that the study of the shock wave dynamics provides a very accurate test of the validity of the GHD equations. 

\begin{figure}[t]
\includegraphics[width=8cm]{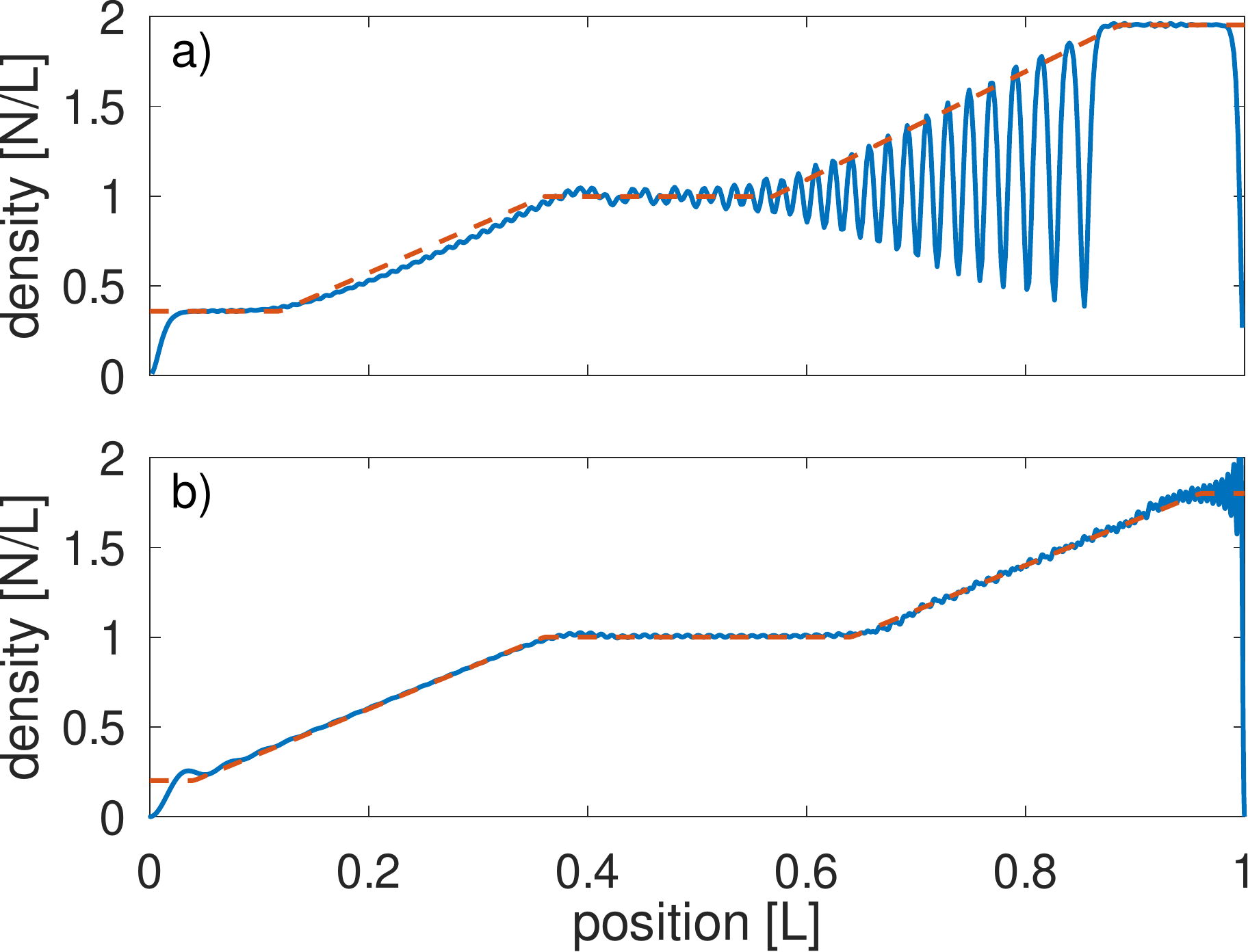}
\caption{\label{fig:GHD_profiles}
(Color online) Comparison of the density profiles for a quench of $0.8 c(\gamma)$ at $t=0.2\times L/c(\gamma)$ predicted by GHD (dashed orange lines) to the a) GP and b) TG results (solid blue lines). For the GHD calculation  we have taken $\gamma=0.01$ for panel a), and  $\gamma=940$ for b).}
\end{figure}
Finally we compare in Fig.~\ref{fig:GHD_profiles} the density profiles at short time $t=0.2\times L/c$ obtained from the microscopic GP and TG calculations to the GHD long wavelength prediction (at $\gamma=0.01$ and $\gamma=940$ respectively). We observe that the GHD prediction reproduces remarkably well the profiles both at small and large interactions: in particular the rarefaction wave is well captured and the GHD reproduces the (upper) envelope of the soliton train in the GP model. This is a strong evidence that indeed GHD captures correctly the DSW dynamics.

\section{Oscillations of the current at finite temperature}
\label{app:finiteT}
\begin{figure}[t]
\includegraphics[width=.95\linewidth]{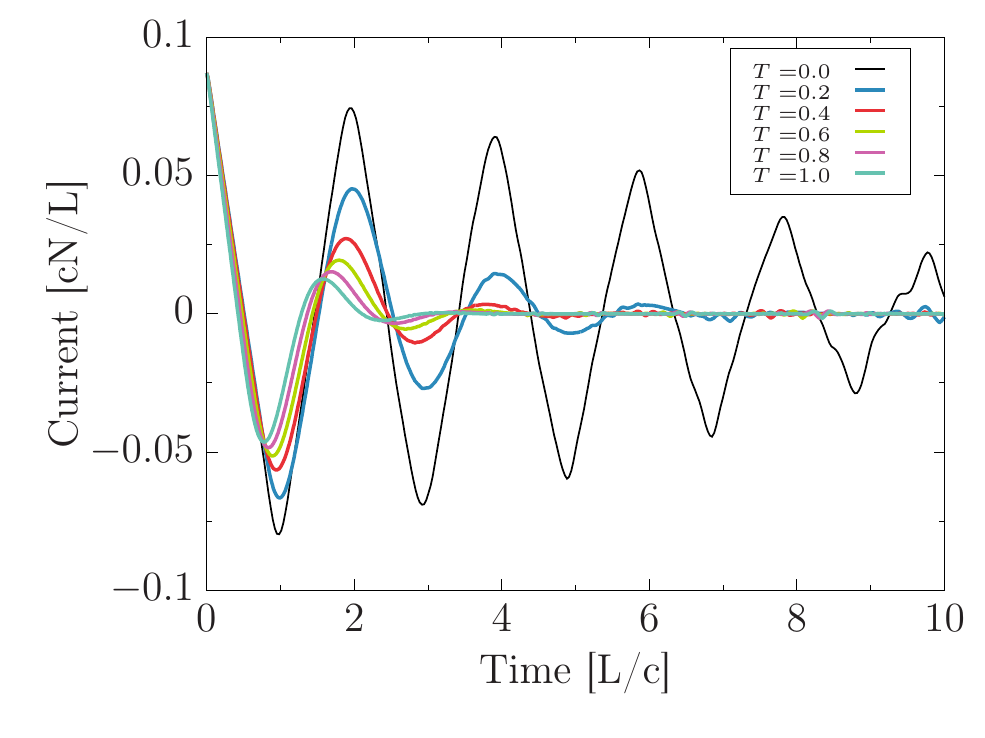}
\caption{
\label{fig:TG_current_temperature} 
Current per particle as a function of time for different temperatures using the exact TG  for $N=23$. Temperatures are given in units of the Fermi temperature and the system is quenched with a velocity boost of $0.087c$. Time is given in units of $L/c(T)$ where the speed of sound $c(T)$ depends on the temperature.
}
\end{figure}

\paragraph{Strong interactions}
In the strongly interacting regime, at finite temperatures, bosonic particles can be described using the Bose-Fermi mapping, in which particles populate the eigenstates of the system following the Fermi-Dirac distribution. When considering a quench into such Fermi sphere, the Hilbert space over which the quenched state projects increases, leading to more low energy excitations during the quench~\cite{Polo_2018}. In \fref{fig:TG_current_temperature} we calculate the total current, \eref{eq:current_temp_TG}, at different temperatures. Note that at temperatures lower than  the Fermi temperature, the current oscillations are still visible and follow a few full oscillations, which shows the robustness of the universal features discussed in the main text. For temperatures of the order of the Fermi temperature, the damping increases dramatically and shock waves diffuse rapidly.

\paragraph{Generalized hydrodynamics}
In order to include finite temperature effects in the GHD we use the Thermodynamic Bethe-Ansatz~\cite{Yang1969}. The initial equilibrium occupation function is:
\[
\bar{n}(k)=\frac{1}{1+e^{\beta\epsilon_k}},
\]
where the pseudo-energy $\epsilon_k$ is the solution of:
\[
\beta\epsilon_k=\beta\left(\frac{\hbar^2k^2}{2m}-\mu\right)-\int\frac{dk^\prime}{2\pi}\phi(k-k^\prime)\ln{\left(1+e^{-\beta\epsilon_{k^\prime}}\right)}.
\]
The box boundary conditions and quench protocol are implemented in the initial state as in the zero temperature case and we use the iterative integration algorithm explained above. 

\begin{figure}[t]
\includegraphics[width=.95\linewidth]{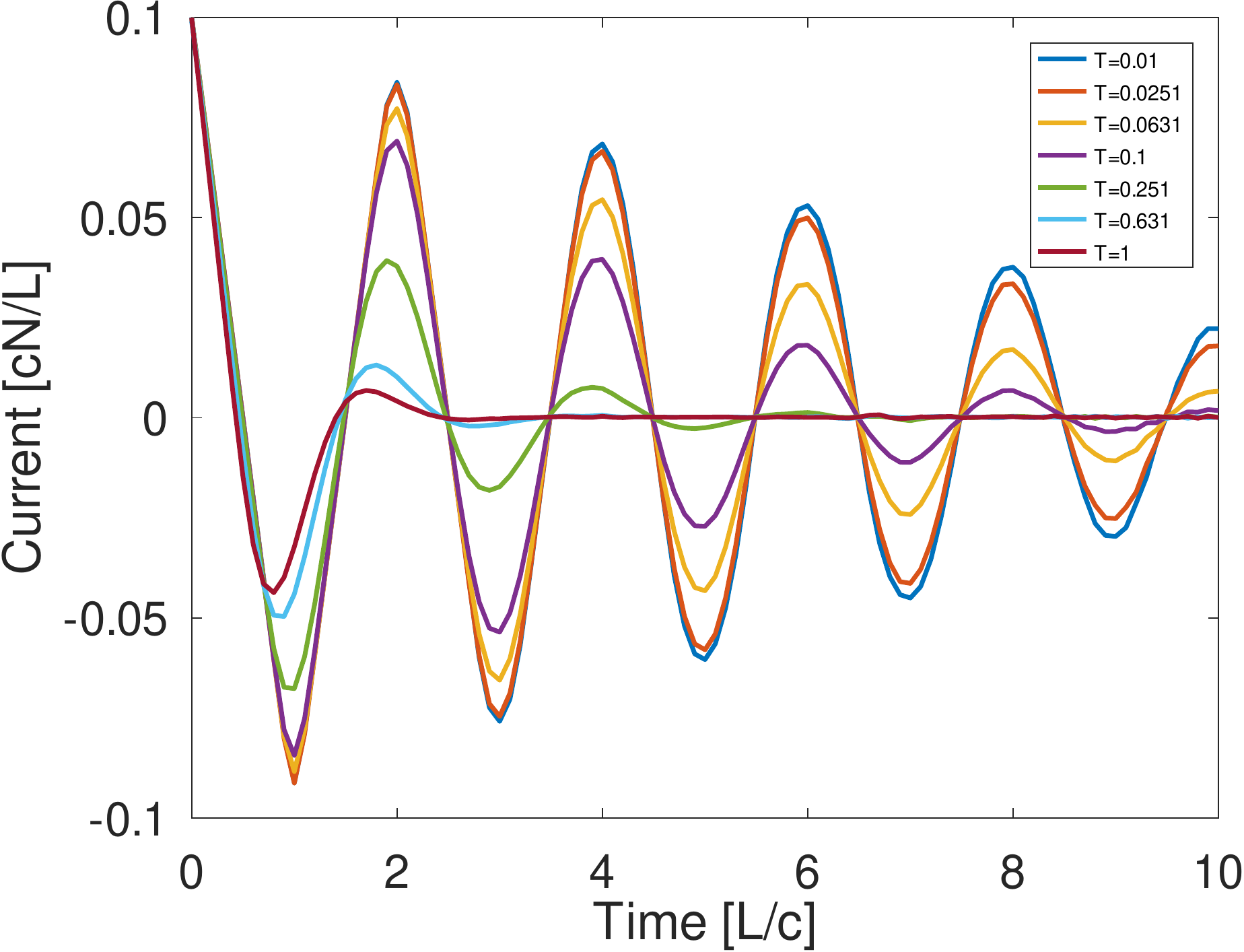}
\caption{
\label{fig:GHD_current_temperature} 
Current (in units of $cN/L$) as a function of time (in units of $L/c(T)$), computed with the GHD approach at $\gamma=1$ and several temperatures covering the range $[0.01,1.3]\times \mu$. For each temperature the initial velocity boost is $0.1c(T)$, where the speed of sound $c(T)$ weakly depends on the temperature.
}
\end{figure}
Figure~\ref{fig:GHD_current_temperature} shows the decay of the current oscillations as temperature increases. The phenomenon reported in the paper is robust up to $T\sim0.25\times\hbar^2n_0^2/(mk_B)$ where $n_0=N/L$ is the one-dimensional density.
 
\section{Identification of the density dips as a soliton train}
\begin{figure}[t]
\includegraphics[width=.95\linewidth]{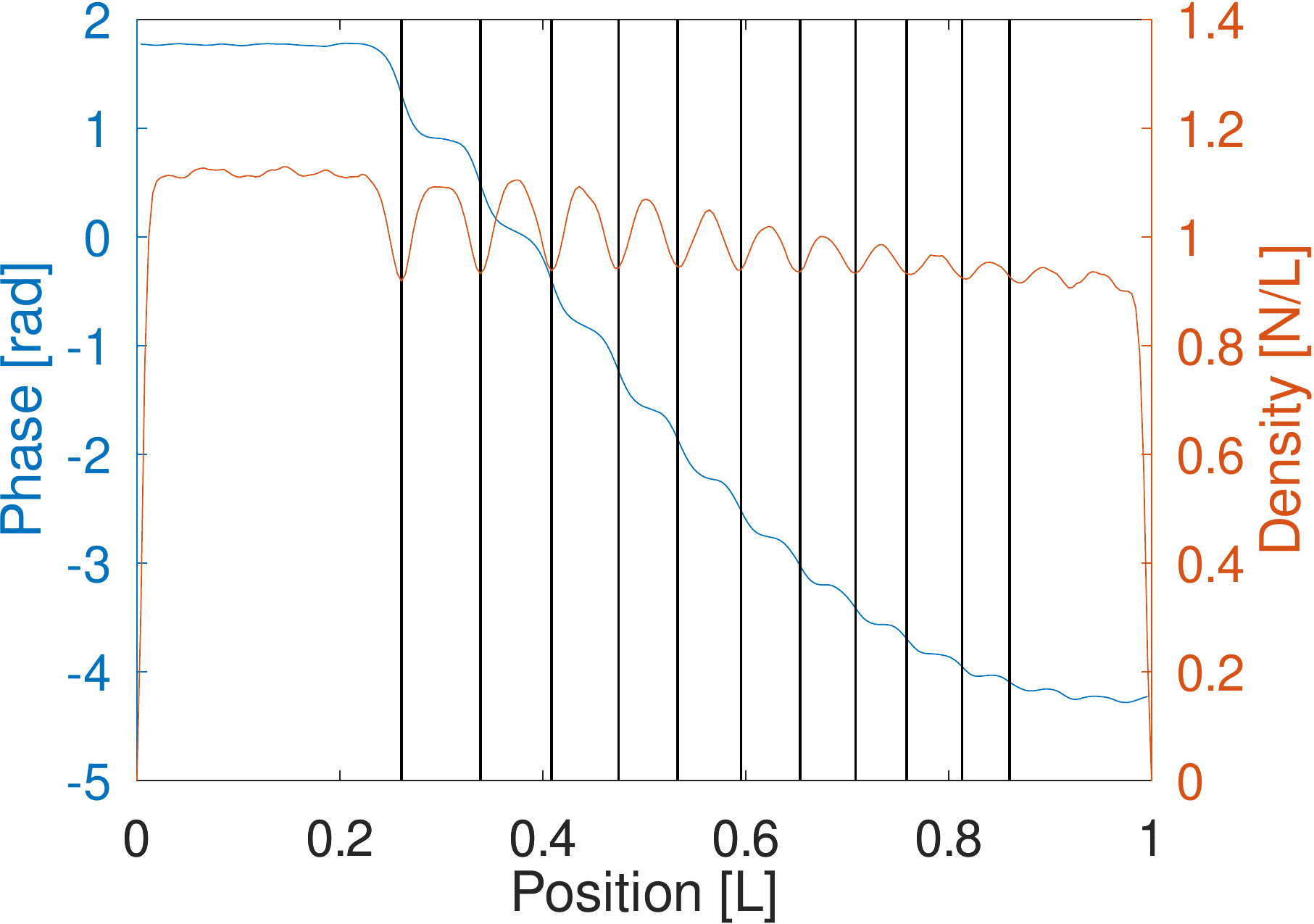}
\caption{
\label{fig:density_dips} 
Density and phase profiles in the GPE simulation at $t=3.26\times L/c_{\rm GP}$. The black vertical lines indicate the points where the slope of the phase is (locally) minimal, and match the position of the density dips.
}
\end{figure}
Figure~\ref{fig:density_dips} shows the density and phase a profile obtained in the GPE equation at an intermediate time $t=3.26\times L/c_{\rm GP}$, where a train of 11 to 13 solitons is seen as small density dips associated to well defined ``steps" in the phase profile. Therefore it seems that for our scenario the density oscillations associated to the shock front propagation are mainly due to fast grey solitons. As the solitons propagate with slightly different speeds (the shallower the faster) and bounce back on the hard wall boundaries, the phase profile can be complicated to interpret at later times where solitons propagates in both directions and overlap. We have checked that the number of generated solitons increases with the quench amplitude.

\section{Density profiles at later times}
\begin{figure}[t]
\includegraphics[width=.95\linewidth]{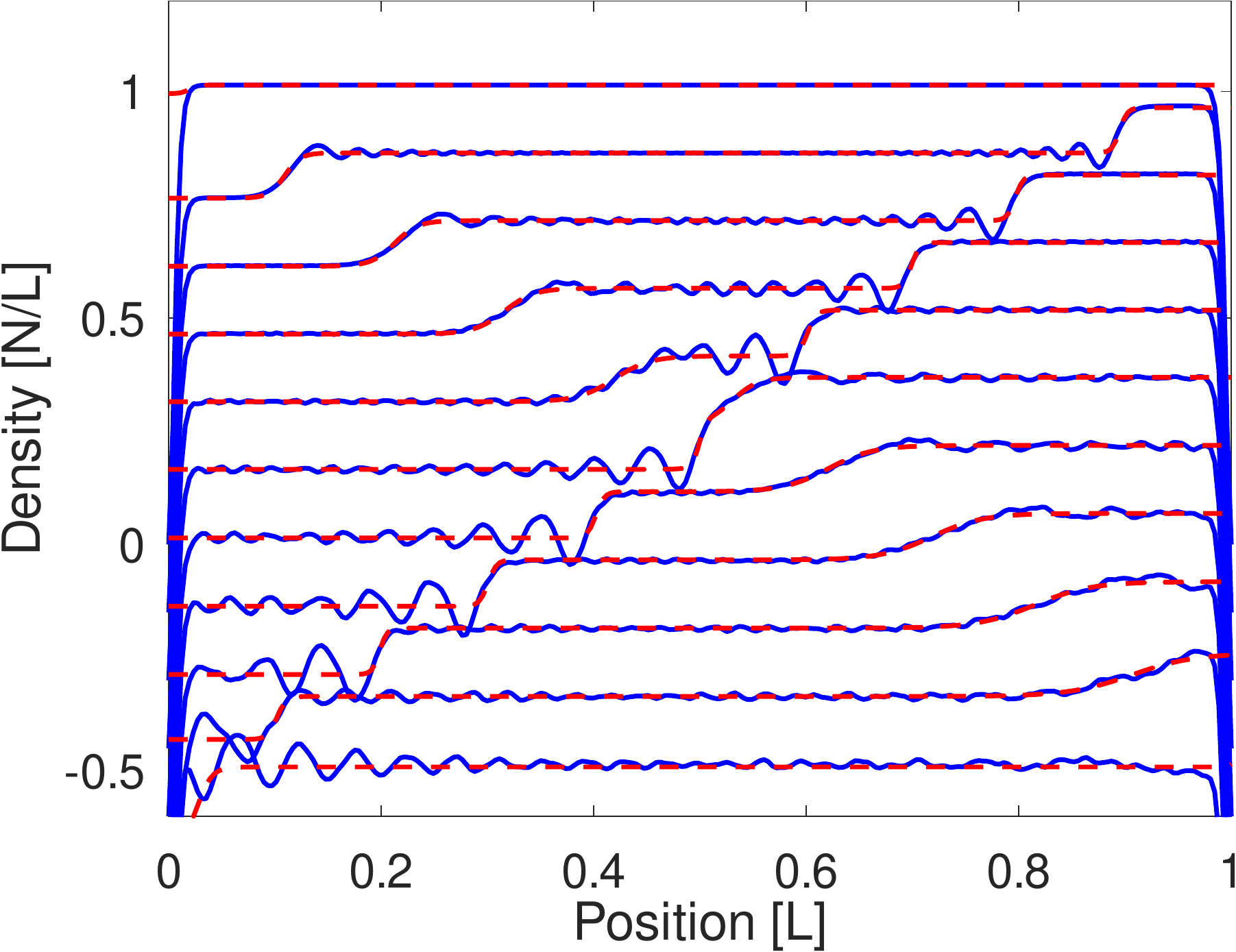}\\ \includegraphics[width=.95\linewidth]{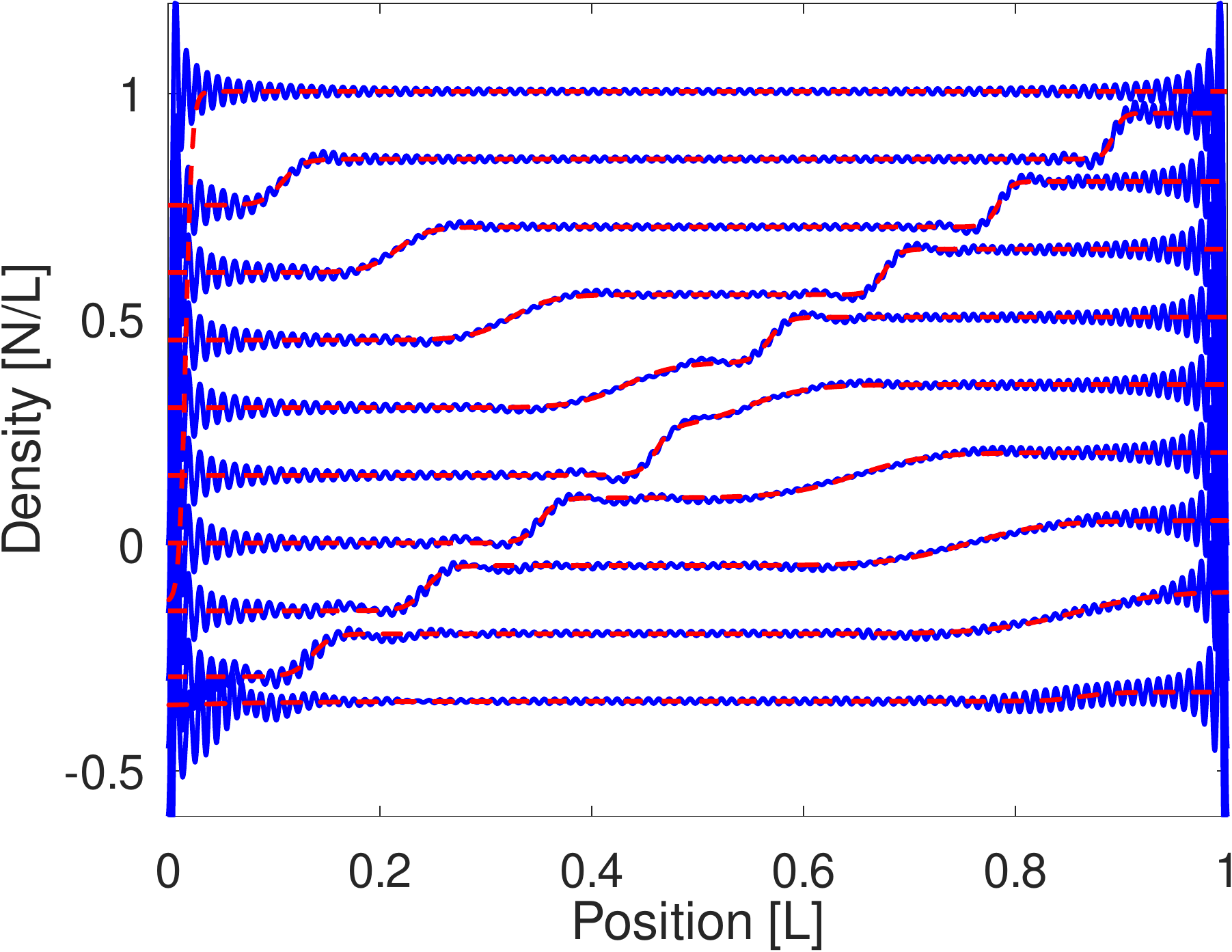}
\caption{
\label{fig:density_model_GPE_TG} 
Density profile in the GPE (top panel) and TG (bottom panel) simulations (solid blue lines), at times: $t=L/c(\gamma)\times\{0,0.1,0.2,0.3,0.4,0.5,0.6,0.7,0.8,0.9,1.0\}$ (from top to bottom). Densities at successive times are vertically shifted by $0.15N/L$ for clarity.
}
\end{figure}

As shown in Fig.~\ref{fig:density_model_GPE_TG} the two propagating fronts meet at $t=0.5\times L/c(\gamma)$ and pass through each other, emphasizing the similarity between the weakly and strongly interacting exact results.

\end{document}